 \RequirePackage{lineno}

\documentclass[aps,prl,twocolumn,amsmath,amssymb,superscriptaddress,showpacs]{revtex4}

\usepackage{graphicx}
\usepackage{dcolumn}
\usepackage{bm}
\usepackage[abs]{overpic} 
\preprint{draftV3.0}

   \newcommand{\ppbar}{\ensuremath{\bar{p}p}}                                                                                             %
   \newcommand{\Bd}{\ensuremath{B^{0}}}                                                                                                   %
   \newcommand{\aBd}{\ensuremath{\overline{B}^{0}}}                                                                            %
   \newcommand{\Bu}{\ensuremath{B^{+}}}                                                                                                   %
   \newcommand{\Bs}{\ensuremath{B_{s}^{0}}}                                                                                               %

   \newcommand{\Bdpipi}{\ensuremath{\Bd\rightarrow\pi^{+}\pi^{-}}} 	
   
   \newcommand{\BdKpi}{\ensuremath{\Bd \rightarrow K^+ \pi^-}}
   \newcommand{\aBdKpi}{\ensuremath{\aBd \rightarrow K^- \pi^+}}
   \newcommand{\BsKpi}{\ensuremath{\Bs \rightarrow K^- \pi^+}}
   \newcommand{\aBsKpi}{\ensuremath{\aBs\rightarrow K^+ \pi^-}}
   \newcommand{\BsKK}{\ensuremath{\Bs \rightarrow  K^+ K^-}}

   \newcommand{\Lppi}{\ensuremath{\Lambda \rightarrow p\pi^{-}}}
   
   \newcommand{\Lbppi}{\ensuremath{\Lambda_{b}^{0} \rightarrow p\pi^{-}}}
   
   \newcommand{\LbpK}{\ensuremath{\Lambda_{b}^{0} \rightarrow pK^{-}}}

   \newcommand{\DKpi}{\ensuremath{D^0\rightarrow K^-\pi^+}}
   \newcommand{\aDKpi}{\ensuremath{\overline{D}^0\rightarrow K^+\pi^-}}             
   
   \newcommand{\Lb}{\ensuremath{\Lambda^0_b}}           
   
   \newcommand{\BsKKsuBdKpidef}{\ensuremath{\frac{f_s}{f_d}\frac{\BR(\Bs\rightarrow K^+K^-)}{\BR(\Bd\rightarrow K^+\pi^-)}}}

   \newcommand{\BdpipisuBdKpidef}{\ensuremath{\frac{\BR(\Bd\rightarrow \pi^+\pi^-)}{\BR(\Bd\rightarrow K^+\pi^-)}}}

   \newcommand{\BR}{\ensuremath{\mathcal B}}                                                                                   %
   \newcommand{\Bhh}{\ensuremath{B \rightarrow h^{+}h'^{-}}}                                                                   %
   \newcommand{\Dhh}{\ensuremath{D^0 \rightarrow h^{+}h'^{-}}}                                                                   %
   \newcommand{\dedx}{\ensuremath{\mathit{dE/dx}}}                                                                              %
   \newcommand{\micron}{\ensuremath{\mu\mathrm{m}}}
   \newcommand{\pgev}{\ensuremath{\mathrm{GeV}/c}}
   \newcommand{\ptot}{\ensuremath{p_{\mathit{tot}}}}   
   \newcommand{\massgev}{\ensuremath{\mathrm{GeV}/c^{2}}}

\newcommand{\lumifb}{ fb$^{-1}$}

\newcommand{\mpipi}{\ensuremath{m_{\pipi}}}
\newcommand{\pipi}{\ensuremath{\pi\pi}}
\newcommand{\stat}{\ensuremath{\mathrm{~(stat)}}}		
\newcommand{\syst}{\ensuremath{\mathrm{~(syst)}}}		
\newcommand{\aBs}{\ensuremath{\overline{B}_{s}^{0}}}
\newcommand{\Lxy}{\ensuremath{L_{T}}}			

\newcommand{\CP}{\ensuremath{\mathit{CP}}}
\newcommand{\acp}{\ensuremath{A_{\CP}}}
\newcommand{\acpraw}{\ensuremath{\tilde{A}_{\CP}}}

\begin{document}
\title{Measurements of Direct \CP\ Violating Asymmetries
           in Charmless Decays of Strange Bottom Mesons and Bottom Baryons}

\affiliation{Institute of Physics, Academia Sinica, Taipei, Taiwan 11529, Republic of China} 
\affiliation{Argonne National Laboratory, Argonne, Illinois 60439, USA} 
\affiliation{University of Athens, 157 71 Athens, Greece} 
\affiliation{Institut de Fisica d'Altes Energies, ICREA, Universitat Autonoma de Barcelona, E-08193, Bellaterra (Barcelona), Spain} 
\affiliation{Baylor University, Waco, Texas 76798, USA} 
\affiliation{Istituto Nazionale di Fisica Nucleare Bologna, $^z$University of Bologna, I-40127 Bologna, Italy} 
\affiliation{University of California, Davis, Davis, California 95616, USA} 
\affiliation{University of California, Los Angeles, Los Angeles, California 90024, USA} 
\affiliation{Instituto de Fisica de Cantabria, CSIC-University of Cantabria, 39005 Santander, Spain} 
\affiliation{Carnegie Mellon University, Pittsburgh, Pennsylvania 15213, USA} 
\affiliation{Enrico Fermi Institute, University of Chicago, Chicago, Illinois 60637, USA}
\affiliation{Comenius University, 842 48 Bratislava, Slovakia; Institute of Experimental Physics, 040 01 Kosice, Slovakia} 
\affiliation{Joint Institute for Nuclear Research, RU-141980 Dubna, Russia} 
\affiliation{Duke University, Durham, North Carolina 27708, USA} 
\affiliation{Fermi National Accelerator Laboratory, Batavia, Illinois 60510, USA} 
\affiliation{University of Florida, Gainesville, Florida 32611, USA} 
\affiliation{Laboratori Nazionali di Frascati, Istituto Nazionale di Fisica Nucleare, I-00044 Frascati, Italy} 
\affiliation{University of Geneva, CH-1211 Geneva 4, Switzerland} 
\affiliation{Glasgow University, Glasgow G12 8QQ, United Kingdom} 
\affiliation{Harvard University, Cambridge, Massachusetts 02138, USA} 
\affiliation{Division of High Energy Physics, Department of Physics, University of Helsinki and Helsinki Institute of Physics, FIN-00014, Helsinki, Finland} 
\affiliation{University of Illinois, Urbana, Illinois 61801, USA} 
\affiliation{The Johns Hopkins University, Baltimore, Maryland 21218, USA} 
\affiliation{Institut f\"{u}r Experimentelle Kernphysik, Karlsruhe Institute of Technology, D-76131 Karlsruhe, Germany} 
\affiliation{Center for High Energy Physics: Kyungpook National University, Daegu 702-701, Korea; Seoul National University, Seoul 151-742, Korea; Sungkyunkwan University, Suwon 440-746, Korea; Korea Institute of Science and Technology Information, Daejeon 305-806, Korea; Chonnam National University, Gwangju 500-757, Korea; Chonbuk National University, Jeonju 561-756, Korea} 
\affiliation{Ernest Orlando Lawrence Berkeley National Laboratory, Berkeley, California 94720, USA} 
\affiliation{University of Liverpool, Liverpool L69 7ZE, United Kingdom} 
\affiliation{University College London, London WC1E 6BT, United Kingdom} 
\affiliation{Centro de Investigaciones Energeticas Medioambientales y Tecnologicas, E-28040 Madrid, Spain} 
\affiliation{Massachusetts Institute of Technology, Cambridge, Massachusetts 02139, USA} 
\affiliation{Institute of Particle Physics: McGill University, Montr\'{e}al, Qu\'{e}bec, Canada H3A~2T8; Simon Fraser University, Burnaby, British Columbia, Canada V5A~1S6; University of Toronto, Toronto, Ontario, Canada M5S~1A7; and TRIUMF, Vancouver, British Columbia, Canada V6T~2A3} 
\affiliation{University of Michigan, Ann Arbor, Michigan 48109, USA} 
\affiliation{Michigan State University, East Lansing, Michigan 48824, USA}
\affiliation{Institution for Theoretical and Experimental Physics, ITEP, Moscow 117259, Russia}
\affiliation{University of New Mexico, Albuquerque, New Mexico 87131, USA} 
\affiliation{Northwestern University, Evanston, Illinois 60208, USA} 
\affiliation{The Ohio State University, Columbus, Ohio 43210, USA} 
\affiliation{Okayama University, Okayama 700-8530, Japan} 
\affiliation{Osaka City University, Osaka 588, Japan} 
\affiliation{University of Oxford, Oxford OX1 3RH, United Kingdom} 
\affiliation{Istituto Nazionale di Fisica Nucleare, Sezione di Padova-Trento, $^{aa}$University of Padova, I-35131 Padova, Italy} 
\affiliation{LPNHE, Universite Pierre et Marie Curie/IN2P3-CNRS, UMR7585, Paris, F-75252 France} 
\affiliation{University of Pennsylvania, Philadelphia, Pennsylvania 19104, USA}
\affiliation{Istituto Nazionale di Fisica Nucleare Pisa, $^{bb}$University of Pisa, $^{cc}$University of Siena and $^{dd}$Scuola Normale Superiore, I-56127 Pisa, Italy} 
\affiliation{University of Pittsburgh, Pittsburgh, Pennsylvania 15260, USA} 
\affiliation{Purdue University, West Lafayette, Indiana 47907, USA} 
\affiliation{University of Rochester, Rochester, New York 14627, USA} 
\affiliation{The Rockefeller University, New York, New York 10065, USA} 
\affiliation{Istituto Nazionale di Fisica Nucleare, Sezione di Roma 1, $^{ee}$Sapienza Universit\`{a} di Roma, I-00185 Roma, Italy} 

\affiliation{Rutgers University, Piscataway, New Jersey 08855, USA} 
\affiliation{Texas A\&M University, College Station, Texas 77843, USA} 
\affiliation{Istituto Nazionale di Fisica Nucleare Trieste/Udine, I-34100 Trieste, $^{ff}$University of Trieste/Udine, I-33100 Udine, Italy} 
\affiliation{University of Tsukuba, Tsukuba, Ibaraki 305, Japan} 
\affiliation{Tufts University, Medford, Massachusetts 02155, USA} 
\affiliation{University of Virginia, Charlottesville, VA  22906, USA}
\affiliation{Waseda University, Tokyo 169, Japan} 
\affiliation{Wayne State University, Detroit, Michigan 48201, USA} 
\affiliation{University of Wisconsin, Madison, Wisconsin 53706, USA} 
\affiliation{Yale University, New Haven, Connecticut 06520, USA} 
\author{T.~Aaltonen}
\affiliation{Division of High Energy Physics, Department of Physics, University of Helsinki and Helsinki Institute of Physics, FIN-00014, Helsinki, Finland}
\author{B.~\'{A}lvarez~Gonz\'{a}lez$^v$}
\affiliation{Instituto de Fisica de Cantabria, CSIC-University of Cantabria, 39005 Santander, Spain}
\author{S.~Amerio}
\affiliation{Istituto Nazionale di Fisica Nucleare, Sezione di Padova-Trento, $^{aa}$University of Padova, I-35131 Padova, Italy} 

\author{D.~Amidei}
\affiliation{University of Michigan, Ann Arbor, Michigan 48109, USA}
\author{A.~Anastassov}
\affiliation{Northwestern University, Evanston, Illinois 60208, USA}
\author{A.~Annovi}
\affiliation{Laboratori Nazionali di Frascati, Istituto Nazionale di Fisica Nucleare, I-00044 Frascati, Italy}
\author{J.~Antos}
\affiliation{Comenius University, 842 48 Bratislava, Slovakia; Institute of Experimental Physics, 040 01 Kosice, Slovakia}
\author{G.~Apollinari}
\affiliation{Fermi National Accelerator Laboratory, Batavia, Illinois 60510, USA}
\author{J.A.~Appel}
\affiliation{Fermi National Accelerator Laboratory, Batavia, Illinois 60510, USA}
\author{A.~Apresyan}
\affiliation{Purdue University, West Lafayette, Indiana 47907, USA}
% ...
\collaboration{CDF Collaboration\footnote{With visitors from $^a$University of Massachusetts Amherst, Amherst, Massachusetts 01003,
$^b$Istituto Nazionale di Fisica Nucleare, Sezione di Cagliari, 09042 Monserrato (Cagliari), Italy,
$^c$University of California Irvine, Irvine, CA  92697, 
$^d$University of California Santa Barbara, Santa Barbara, CA 93106
$^e$University of California Santa Cruz, Santa Cruz, CA  95064,
$^f$CERN,CH-1211 Geneva, Switzerland,
$^g$Cornell University, Ithaca, NY  14853, 
$^h$University of Cyprus, Nicosia CY-1678, Cyprus, 
$^i$University College Dublin, Dublin 4, Ireland,
$^j$University of Fukui, Fukui City, Fukui Prefecture, Japan 910-0017,
$^k$Universidad Iberoamericana, Mexico D.F., Mexico,
$^l$Iowa State University, Ames, IA  50011,
$^m$University of Iowa, Iowa City, IA  52242,
$^n$Kinki University, Higashi-Osaka City, Japan 577-8502,
$^o$Kansas State University, Manhattan, KS 66506,
$^p$University of Manchester, Manchester M13 9PL, England,
$^q$Queen Mary, University of London, London, E1 4NS, England,
$^r$Muons, Inc., Batavia, IL 60510,
$^s$Nagasaki Institute of Applied Science, Nagasaki, Japan, 
$^t$National Research Nuclear University, Moscow, Russia,
$^u$University of Notre Dame, Notre Dame, IN 46556,
$^v$Universidad de Oviedo, E-33007 Oviedo, Spain, 
$^w$Texas Tech University, Lubbock, TX  79609, 
$^x$Universidad Tecnica Federico Santa Maria, 110v Valparaiso, Chile,
$^y$Yarmouk University, Irbid 211-63, Jordan,
$^{gg}$On leave from J.~Stefan Institute, Ljubljana, Slovenia, 
}}
\noaffiliation

\date{February 2, 2011}
\begin{abstract}
We report measurements of direct \CP--violating asymmetries in
charmless decays of neutral bottom hadrons to pairs of charged hadrons
with the upgraded Collider Detector at the Fermilab Tevatron.  Using a
data sample corresponding to \mbox{1\lumifb} of integrated luminosity,
we obtain the first measurements of direct \CP\ violation in bottom 
strange mesons, $\acp(\BsKpi) = +0.39 \pm 0.15\stat \pm 0.08\syst$,
and bottom baryons, $\acp(\Lbppi) = +0.03 \pm 0.17\stat \pm 0.05\syst$
and $\acp(\LbpK) = +0.37 \pm 0.17\stat \pm 0.03\syst$.  
In addition, we measure \CP\ violation in \BdKpi\ decays with  $3.5\sigma$ significance, $\acp(\BdKpi) = -0.086\pm 0.023 \stat \pm 0.009\syst$, in agreement with the current world average.
Measurements of branching fractions of \BsKK\ and \Bdpipi\ decays are also updated.
\end{abstract}

\pacs{13.25.Hw, 11.30.Er, 14.20.Mr, 14.40.Nd}
\maketitle

Noninvariance of the fundamental interactions under the combined
symmetry transformation of charge conjugation and parity inversion (\CP\ violation) 
is an established experimental fact. The vast majority of experimental 
data are well described by the standard model (SM), 
and have supported the success of the Cabibbo-Kobayashi-Maskawa (CKM)~\cite{Kobayashi:1973fv}
theory of quark-flavor dynamics.
However, additional sources of \CP\ violation are required to explain the matter--anti-matter asymmetry of the Universe in standard big-bang cosmology.
This would have profound consequences on our understanding of fundamental interactions.

Violation of \CP\ is \emph{direct} if the partial decay-width~($\Gamma$) of a
particle into a final state differs from the width of the corresponding
antiparticle into the \CP-conjugate final state.
In recent times, the pattern of direct \CP\ violation in
charmless mesonic decays of $B$ mesons has shown some 
unanticipated discrepancies from expectations. 
Under standard assumptions of isospin symmetry and smallness of contributions 
from higher-order processes, 
similar \CP\ asymmetries are predicted for \BdKpi\ and $\Bu\to K^+\pi^0$ decays~\cite{Keum:2002vi,Beneke:2003zv}. 
However, experimental data show a significant discrepancy \cite{acp_bfactories}, 
which has prompted intense experimental and theoretical
activity. Several simple extensions of the standard model
could accommodate the discrepancy~\cite{newphysics}, but  
uncertainty on the contribution of higher-order SM amplitudes 
has prevented a firm conclusion~\cite{allstandard}. 
The violation of \CP\ symmetry in charmless modes remains, therefore, a very interesting subject of study.
Rich samples of bottom-flavored hadrons of all 
types from the Tevatron offer the opportunity to explore 
new territory in the field of \Bs\ mesons and $b$-flavored baryons.
Additional information coming from different decays 
yields further constraints on the possible 
explanations of previous findings, and may possibly reveal new 
deviations from expectations. 

Specifically, measurements of direct \CP\ violation in  \BsKpi\ 
decays have been proposed as a nearly model-independent test
for the presence of non-SM physics~\cite{Gronau:2000md,Lipkin:2005pb}. 
The relationships between charged-current quark 
couplings in the SM predict a well-defined hierarchy between direct 
\CP\ violation in \BdKpi\ and \BsKpi\ decays, yielding a significant asymmetry for the latter, of 
about 40\%. This large effect allows easier experimental 
investigation and any discrepancy may indicate contributions from non-SM amplitudes.

Supplementary information could come from \CP\ violation 
in bottom baryons, an effect which has not been measured so far.
Interest in charmless $b$--baryon decays is prompted by branching fractions recently 
observed being larger than expected \cite{Mohanta:2000nk, Lu:2009cm, Aaltonen:2008hg}. 
Asymmetries up to about 10\% are predicted for \LbpK\ and \Lbppi\ decays 
in the SM~\cite{Mohanta:2000za, Lu:2009cm}, and are accessible with current CDF event samples.


In this letter we report the first measurement of direct \CP\ violation
in decays of bottom strange mesons and bottom baryons. We use 1\lumifb\ of \ppbar\  
collisions at $\sqrt{s} = 1.96$ TeV, collected by the upgraded Collider Detector (CDF II) 
at the Tevatron. The \CP--violating asymmetries are measured in the recently 
established~\cite{Aaltonen:2008hg} \BsKpi, \Lbppi\  and \LbpK\ decays~\cite{C-conjugate}. 
We also update our previous measurements~\cite{Abulencia:2006psa} of asymmetry in the \BdKpi\ decay,  
and branching fractions of \Bdpipi\ and \BsKK\ decays.

The CDF II detector is described in detail in Ref.~\cite{Acosta:2004yw} with the detector sub-systems relevant for this 
analysis discussed in Ref.~\cite{Abulencia:2006psa}.
The data are collected by a three-level trigger system.
At level~1, tracks are reconstructed 
in the transverse plane.
Two opposite-charge  particles are required, with reconstructed transverse momenta $p_{T1}, p_{T2} > 2~\pgev$, 
the scalar sum $p_{T1}+p_{T2} > 5.5~\pgev$, and an azimuthal opening-angle $\Delta\phi < 135^{\circ}$. 
At level~2, 
tracks are combined with silicon hits and their impact parameter $d$ (transverse distance of closest approach to the beam line) 
is determined with 45 \micron\ resolution (including the beam spread) and required to be $0.1 < d < 1.0$ mm. 
A tighter opening-angle requirement, $20^{\circ} < \Delta\phi < 135^{\circ}$, is also applied.
Each track pair is then used to form a $B$ candidate, which is  
required to have an impact parameter $d_B < 140~\micron$ and to have travelled a 
distance $\Lxy > 200~\micron$ in the transverse plane. At level~3,  a cluster of computers confirms the selection with a full event reconstruction. 

\begin{figure}[htb]
\includegraphics[scale=0.35]{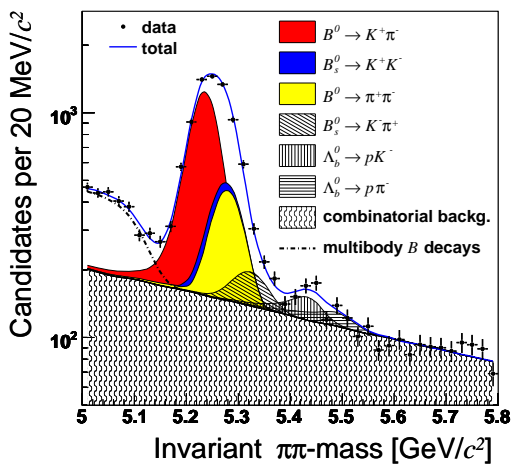}
\caption{\label{fig:projections} Mass distribution
of the 13~502 reconstructed candidates. The charged-pion mass is assigned to both tracks. The total projection and projections of each signal and background component of the likelihood fit are overlaid on  the data distribution. Signals and multi-body $B$ backgrounds are shown stacked on the combinatorial 
background component.}
\end{figure}

The offline selection is based on a more accurate determination of the same quantities 
used in the trigger, with the addition of requirements on two other observables:
the isolation ($I_{B}$) of the $B$ candidate~\cite{Isolation}, and the quality of the three-dimensional fit ($\chi^{2}$ with 1 d.o.f.) 
of the decay vertex of the $B$ candidate \cite{Aaltonen:2008hg}.  
Asymmetries in the rarer \BsKpi\ and \Lb\ decays are measured using the selection 
in Ref.~\cite{Aaltonen:2008hg}. For the measurement of the \BdKpi asymmetry,  instead,  the selection is optimized by minimizing the 
expected variance of the measurement, evaluated by performing the full analysis on a set of simulated 
samples obtained with varied selection criteria~\cite{Morello-Thesis}.
This procedure yields the final selection: $I_B>0.5$, $\chi^{2}<7$, $d > 100~\micron$, 
$d_B < 80~\micron$, and $\Lxy > 300~\micron$. 
Only one $B$ candidate per event is found after this  selection, and a mass ($m_{\pi\pi}$) is assigned to each, 
using a nominal charged-pion mass assignment for both decay products. The resulting mass distribution is shown in 
Fig.~\ref{fig:projections}. 
A large peak is visible, dominated by the overlapping contributions of 
the \BdKpi, \Bdpipi, and \BsKK\ decays~\cite{Abulencia:2006psa}. Signals for \BsKpi, \Lbppi, and \LbpK\ 
modes populate masses higher than the main peak (5.33--5.55 \massgev)~\cite{Aaltonen:2008hg}. 
Backgrounds include mis-reconstructed multi-body $b$--hadron decays (physics background) and random 
pairs of particles (combinatorial background).

We incorporate kinematic and particle identification information in an unbinned likelihood fit~\cite{Abulencia:2006psa,Aaltonen:2008hg} to determine the fraction of each mode and the charge asymmetries, uncorrected for instrumental effects, $\acpraw=[N_{b\to f}-N_{\bar{b}\to \bar{f}}]/[N_{b\to f}+N_{\bar{b}\to \bar{f}}]$ of the flavor-specific decays \BdKpi, \BsKpi, and $\Lambda^0_b\to p\pi^-, pK^-$. For each channel, $N_{b\to f}$ ($N_{\bar{b}\to {\bar{f}}}$) is the reconstructed number of decays of hadrons containing the $b$  ($\bar{b}$) quark into the final state $f$ ($\bar{f}$). 
The decay flavor is inferred from the charges of final state particles assuming equal numbers of $b$ and $\bar{b}$ quarks at production (dominated by the strong interaction).  Any effect from \CP\ violation in $b$--meson flavor mixing is assumed negligible~\cite{PDG}. 

\begin{figure}[htb]
\includegraphics[scale=0.21]{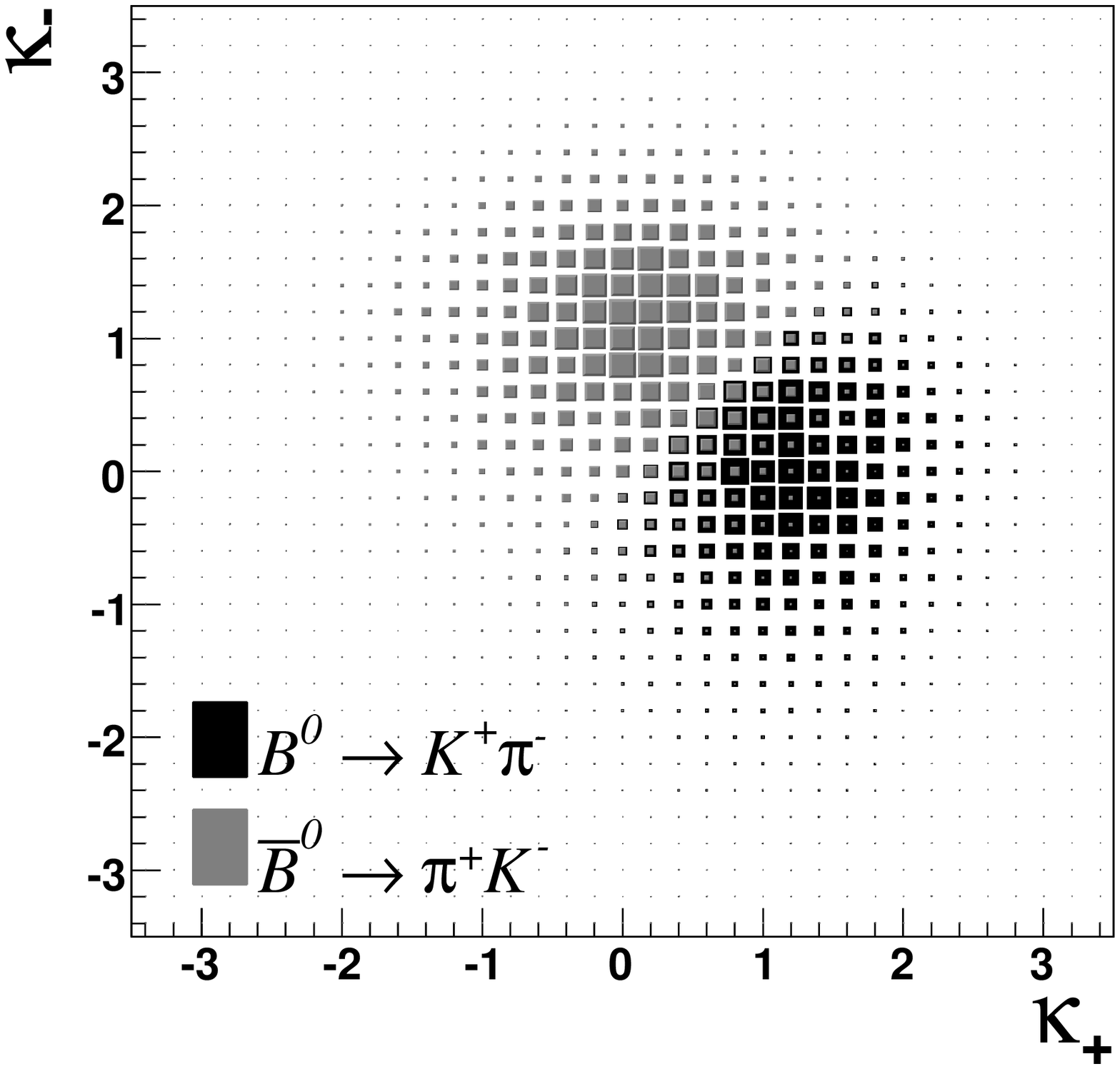}
\includegraphics[scale=0.21]{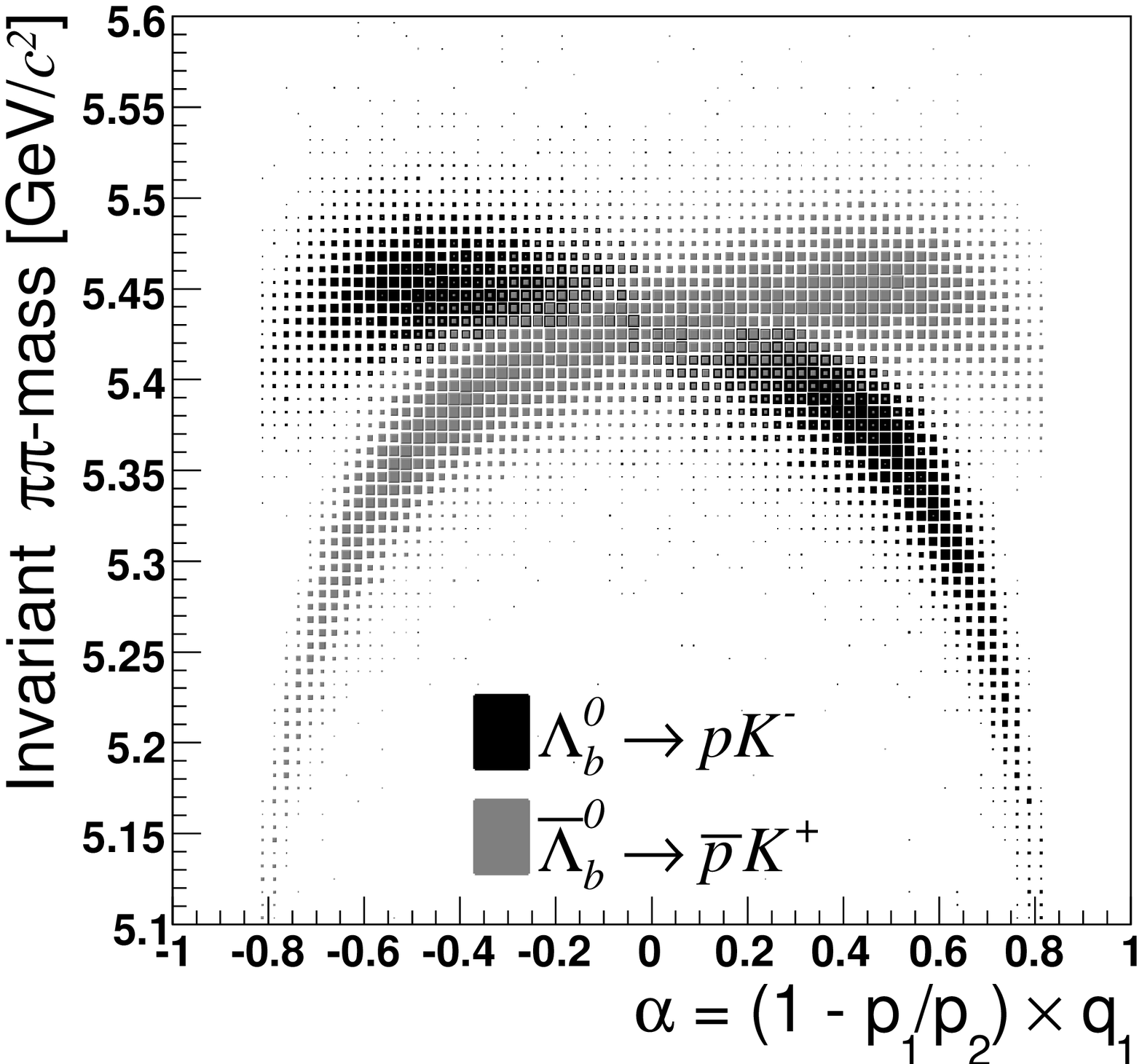}
\includegraphics[scale=0.21]{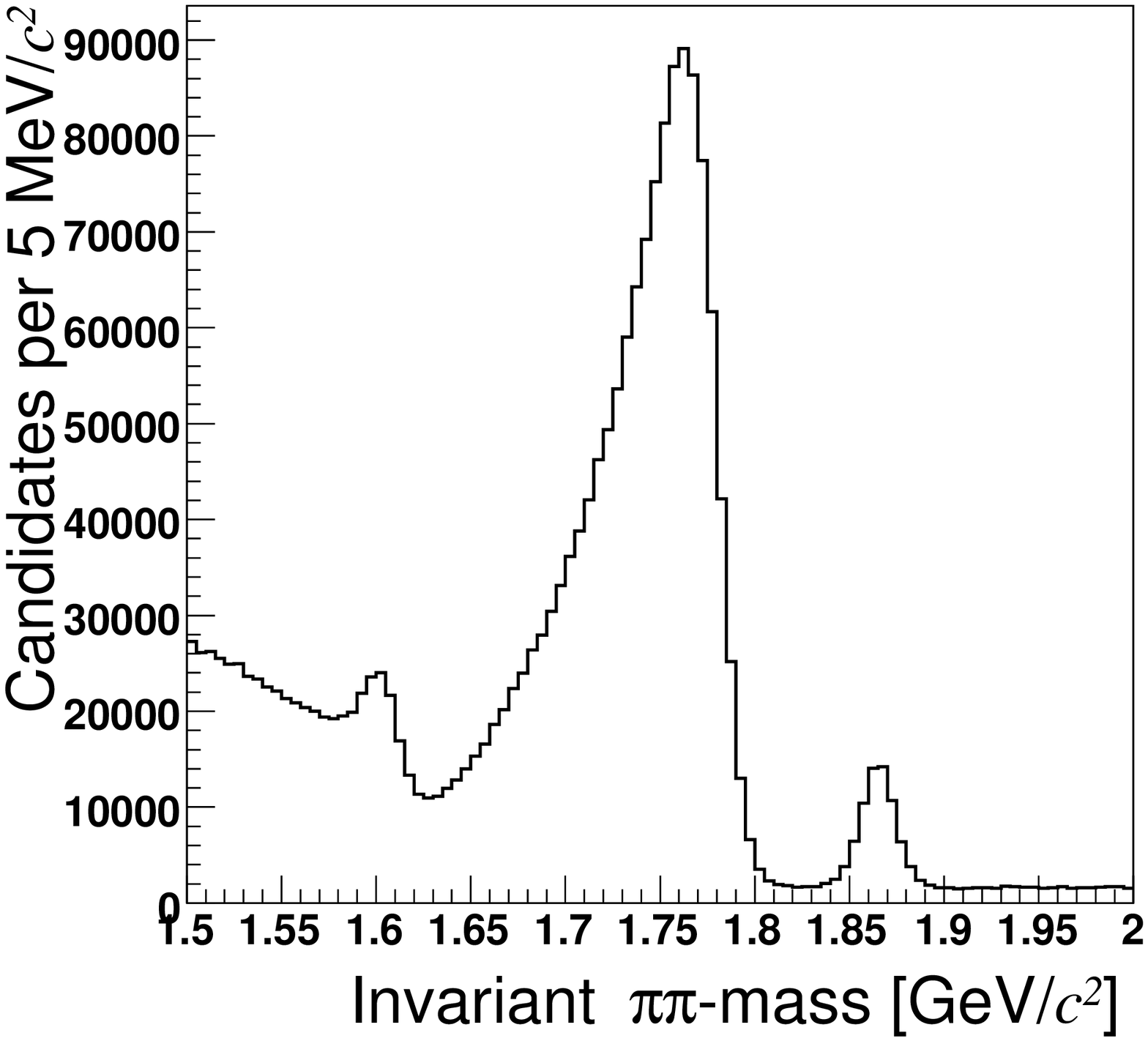}
\includegraphics[scale=0.21]{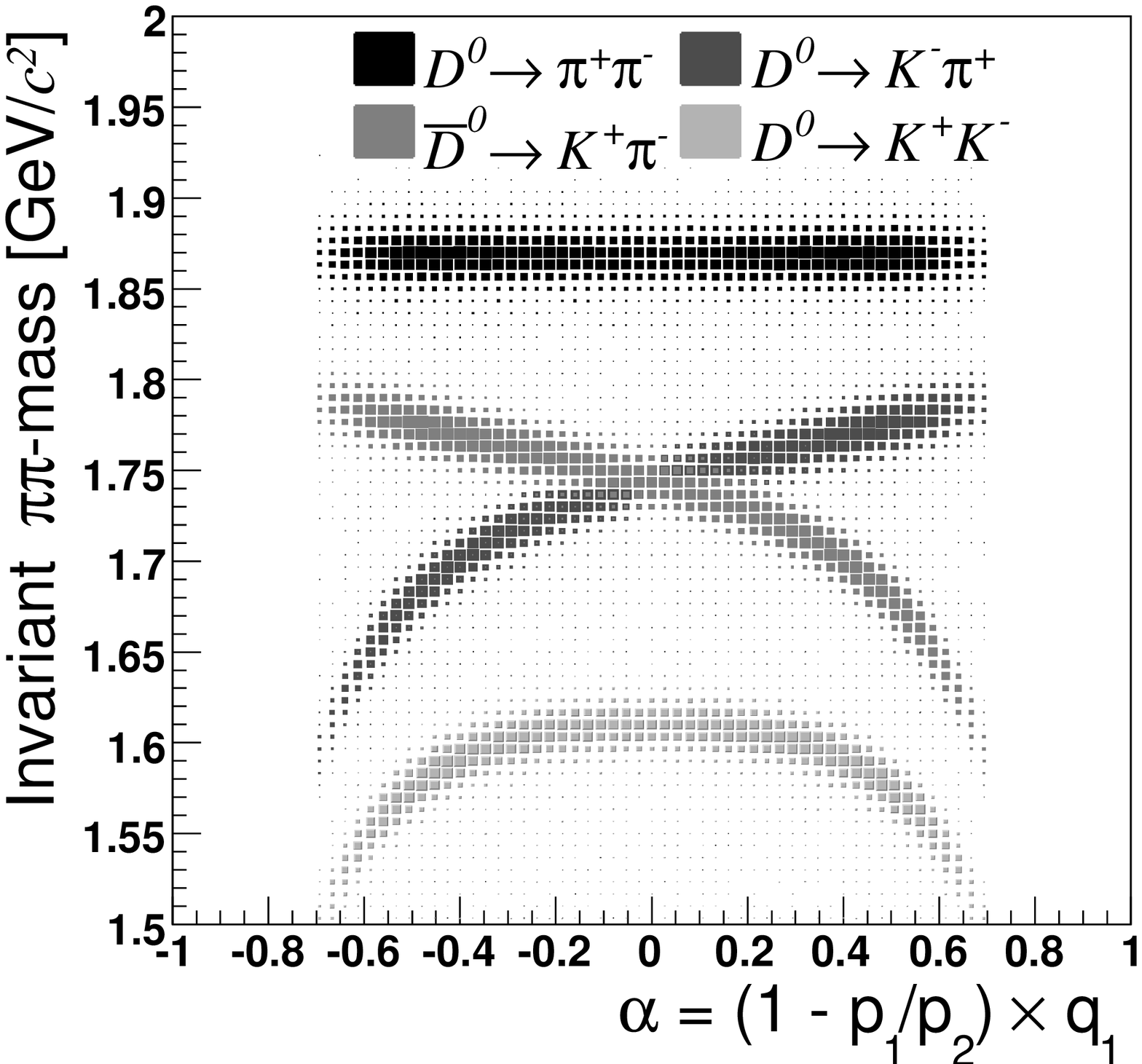}
\caption{Joint kaonness distribution for the positive (abscissa) and negative (ordinate) final state particles
in \BdKpi\ decays as determined from the calibration data of charm decays (top left).
Dipion mass as a function of $\alpha$ for simulated \LbpK\ decays (top right). Mass of \Dhh\ candidates with 
pion assignment to both final state particles (bottom left). Same quantity as a function of $\alpha$ for simulated \Dhh\ decays (bottom right)}
\label{fig:kaonness_pdf}
\end{figure}

The whole kinematic information is summarized by three loosely correlated observables~\cite{Aaltonen:2008hg}: the mass \mpipi; the signed momentum imbalance $\alpha = (1-p_1/p_2) \times q_{1}$, where $p_1$ ($p_2$) is the lower (higher) of the particle momenta, and $q_1$ is the sign of the charge of the particle of momentum $p_{1}$; and the scalar sum of particle momenta  $\ptot=p_1 + p_2$.
Particle identification relies on measurement of the specific ionization (\dedx) 
in the drift chamber. For charged kaons and pions the \dedx\ response was calibrated 
with a sample of $1.5\times 10^6$ $D^{*+}\to D^0\pi^+$ decays, using the charge of the 
pion from $D^{*+}$ decay to identify the products of the Cabibbo--favored $D^0$ decay. 
For protons we used  124~000 $\Lambda\to p \pi^{-}$ decays, 
 where the kinematics and the momentum threshold of the trigger allow unambiguous identification 
of the decay products~\cite{Morello-Thesis,Volpi-Thesis}. 
Identification information for each particle 
is summarized by a single observable in our fit (``kaonness''), 
defined as $\kappa=(dE/dx - dE/dx_{\pi})/(dE/dx_{K} - dE/dx_{\pi})$, where \dedx\ is the 
observed response, and $dE/dx_{\pi(K)}$ is the average responses expected 
for pions (kaons). The separation between $K^+\pi^-$ or $p\pi^-$ 
final states and their charge--conjugates is in excess of 2.1$\sigma$ (Fig.~\ref{fig:kaonness_pdf}). 
Although a lower $dE/dx$ separation is available between $pK^-$ and $\bar{p}K^+$, due to similar 
ionization rates of protons and kaons, sufficient discrimination is achieved from their greater kinematics differences.
The background model allows for independent contributions of positively 
and negatively charged pions, kaons, protons, and electrons, whose fractions 
are determined by the fit.  Muons are indistinguishable from pions with the available 10\% fractional \dedx\
resolution and are therefore incorporated into the pion component.


\begin{table*}[htb]
\caption{\label{tab:ACPs} Raw signal yields determined by the fit and final results.  The first uncertainty is statistical, the second is systematic. Absolute branching fractions are derived by normalizing to the known value ${\mathcal B}(\mbox{\BdKpi})~=~(19.4\pm 0.6) \times 10^{-6}$, and assuming the average value at high energy for the production fraction $f_{s}/f_{d}= 0.282\pm 0.038$~\cite{PDG}.} 
{
\begin{tabular}{lccccccccccccc}
\hline\hline
Mode && $N_{b \to f}$ && $N_{\bar{b} \to \bar{f}}$ &&  $\acp(b\to f)$ (\%) &&   \multicolumn{3}{c}{Relative \BR}   && Absolute \BR ($10^{-6}$)\\
\hline
\BdKpi	&& $1836 \pm 61$  && $2209 \pm 64$ &&   $-8.6 \pm 2.3 \pm 0.9$  && \multicolumn{3}{c}{--}   && --      \\
\BsKpi	&& $160  \pm 26$  && $  70 \pm 22$ &&  $+39  \pm 15  \pm 8 $  && \multicolumn{3}{c}{--}   && --    \\
\LbpK	&& $80   \pm 14$  && $  36 \pm 11$ &&  $+37  \pm 17  \pm 3 $    && \multicolumn{3}{c}{--}   && --      \\
\Lbppi	&& $40   \pm 10$  && $  38 \pm 9$  &&  $+3  \pm 17  \pm 5 $    && \multicolumn{3}{c}{--}   && --       \\
\Bdpipi	&&  \multicolumn{3}{c}{1121 $\pm$ 63} && --  && \BdpipisuBdKpidef    &=&    0.259 $\pm$ 0.017 $\pm$ 0.016  && 5.02 $\pm$ 0.33 $\pm$ 0.35   \\
\BsKK	&&  \multicolumn{3}{c}{1307 $\pm$ 64}&& --   && \BsKKsuBdKpidef      &=&    0.347 $\pm$ 0.020 $\pm$ 0.021  && 23.9 $\pm$ 1.4 $\pm$ 3.6     \\ 
\hline\hline
\end{tabular}
}
\end{table*}



The signal yields from the fit (Table~\ref{tab:ACPs}) 
are corrected for different detection efficiencies to determine the physical asymmetries, $\acp(b\to f)$, defined as 
\begin{equation}
\frac{\BR(b\to f)-  \BR(\bar{b}\to \bar{f}) }
                       {\BR(b\to f)+  \BR(\bar{b}\to \bar{f}) } = \frac{N_{b\to f}-  c_{f} N_{\bar{b}\to \bar{f}} }
                       {N_{b\to f}+  c_{f} N_{\bar{b}\to \bar{f}} },
\label{eq:acp_corr}
\end{equation}

\noindent
where $c_{f} = \varepsilon(f)/\varepsilon{(\bar{f})}$ is the ratio between 
the efficiencies for triggering and reconstructing the final state $f$ 
with respect to the state $\bar{f}$. The $c_{f}$ factors 
correct for detector-induced charge asymmetries, and are extracted 
from control samples in data. Simulation is only used to account for 
small differences between the kinematics of \Bhh\ decays and control signals.
The corrections for $f=K^+\pi^-$ are extracted from a sample of about 
700~000 $D^0\rightarrow K^-\pi^+$ decays, reconstructed in the same data set.
By imposing the same offline selection to the $D^0$ decays we obtain $K^{\mp}\pi^{\pm}$ 
final states in a similar kinematic region as our signals (see Fig.~\ref{fig:kaonness_pdf}). 
We assume that $K^+\pi^-$ and $K^-\pi^+$ final states from charm decays 
are produced in equal numbers at the Tevatron,  because production 
is dominated by the strong interaction and, compared to the detector effects
to be corrected, the possible \CP--violating asymmetry in $D^0\to K^-\pi^+$ 
decays is tiny ($<10^{-3}$) as predicted by the SM \cite{Bianco:2003vb} and 
confirmed by current experimental determinations~\cite{babar_and_belle_acp_d0}. 
We also checked that possible asymmetries in $D^0$ meson yields induced 
by \CP\ violation in $B \to DX$ decays are small and can be neglected~\cite{Morello-Thesis}. 
Therefore, any asymmetry between observed numbers of reconstructed 
$K^-\pi^+$ and $K^+\pi^-$ charm decays can be ascribed to detector-induced 
effects and used to extract the desired correction factors. 
The ratio $N_{\aDKpi}/N_{\DKpi}$ is measured with the same fit used for the signal. 
The \dedx\ information is not used because kinematics alone 
is sufficient to provide an excellent separation in charm decays, as shown in Fig.~\ref{fig:kaonness_pdf}. 
We checked separately that  \dedx\ information does not introduce additional charge asymmetries~\cite{Morello-Thesis}. 
We find $c_{K^{-}\pi^{+}}=0.9871~\pm~0.0027$, which is consistent and more 
precise than a previous estimate based on simulation~\cite{Acosta:2004ts}. 
For the \Lbppi\ asymmetry, the factor $c_{p\pi^{-}}=1.0145 \pm 0.0075$ 
is extracted using a similar strategy applied to a control sample of \Lppi\ decays~\cite{Volpi-Thesis}. 
This factor is dominated by the different interaction probability of protons and 
antiprotons with detector material. In the measurement of \CP\ violation in \LbpK\ decays, 
instrumental charge-asymmetries induced in both kaons and 
protons are relevant. The $c_{pK^{-}}$ factor is extracted by combining
the previous ones and assuming the trigger and reconstruction efficiency for two 
particles factorizes as the product of the single-particle efficiencies.
Corrections are also applied for the branching ratio measurements.
These corrections do not exceed 7\% and account for differences in trigger and reconstruction efficiency between channels due 
to different lifetimes and kinematics (from simulation), and isolation properties (from control samples of 
fully reconstructed $\Bd \to J/\psi K^*(892)^0$ and $\Bs \to J/\psi \phi$ decays).

The dominant contributions to the systematic uncertainties on the asymmetry measurements come from the uncertainty 
on the \dedx\ calibration and parameterization, the uncertainty on the combinatorial background model, and the 
uncertainty on $b$-hadron masses. Smaller contributions are assigned for the uncertainty on the global mass scale and the $c_{f}$ corrections. 
The uncertainty on the \dedx\ model dominates also the systematic uncertainty for the branching ratio measurements, 
for which the mass scale (in the \Bdpipi\ case) and the uncertainty on the difference in isolation efficiency between $\Bd$ and $\Bs$ mesons (\BsKK) also play a role. 
The results  are reported in Table~\ref{tab:ACPs}. 
We report $3.5\sigma$ evidence of \CP\ violation in \BdKpi\ decays. The observed 
asymmetry is consistent, and of comparable accuracy, with current results from 
asymmetric $e^+e^-$ colliders \cite{acp_bfactories}. 
It is also consistent with the result in Ref.~\cite{Abulencia:2006psa} and supersedes it. 
The \BsKpi\ result is the first measurement of direct \CP\ 
violation in bottom strange mesons. It differs by 2.3$\sigma$ 
from zero and it is consistent with recent theoretical predictions~\cite{Beneke:2003zv,theory}.  
It allows the first experimental verification of the model--independent test proposed in 
Ref.~\cite{Lipkin:2005pb}. Under the assumption of equal \Bd\ and \Bs\ lifetimes, 
using the measurement of the \BsKpi\ branching ratio~\cite{Aaltonen:2008hg} and known 
values for the branching ratio and \CP--violating asymmetry in \BdKpi\ decays, and 
the $b$--quark fragmentation probabilities~\cite{PDG}, we obtain 
$R= [\Gamma(\BdKpi) -\Gamma(\aBdKpi)]/[\Gamma(\aBsKpi) -\Gamma(\BsKpi)] = 0.85 \pm 0.42\stat\pm 0.13\syst$,  
which is consistent with the standard prediction, $R^{\mathrm{SM}}=1$~\cite{Lipkin:2005pb}. 
The first measurement of \CP\ violation in bottom baryons is also reported. 
The observed asymmetry in the \LbpK\ decay is 2.1$\sigma$ from zero. 
The \Lbppi\ result is consistent with zero. However, the limited experimental  
precision does not allow a conclusive discrimination between the standard 
model prediction (8\%) and much suppressed values ($\approx 0.3\%$) expected in 
R--parity violating supersymmetric scenarios~\cite{Mohanta:2000za}. 
\par
Table~\ref{tab:ACPs} includes also improved measurements of \BsKK\ and \Bdpipi\ 
\CP--averaged branching fractions, using the \BdKpi\ channel as a reference. 
Results are consistent with previous CDF measurements~\cite{Abulencia:2006psa} 
and supersede them. The \BsKK\ result is the most precise to date and consistent with recent 
theoretical predictions
~\cite{Beneke:2003zv,theory,Sun:2002rn,Buras:2004ub}.
Theory uncertainties, which are significantly larger 
than the experimental ones, prevent sensible discrimination between models. 
The present measurement of $\BR(\Bdpipi)$ agrees with  
measurements at $e^+e^-$ colliders~\cite{Aubert:2006fha} 
with comparable accuracy.
The dominant systematic uncertainties are limited by the finite size of control samples and should 
decrease in future extensions  of the measurements.\par
In conclusion, we have measured \CP--violating asymmetries in charmless \Bd, \Bs, and \Lb\ 
decays into pairs of charged hadrons reconstructed in CDF data. We report the first 
measurement of direct \CP\ violation in bottom strange mesons, the first 
measurement of \CP\ violation in bottom baryons, evidence for \CP\ violation in 
\BdKpi\ decays, and updated measurements of the \BsKK\ and \Bdpipi\ branching fractions.

\begin{acknowledgments}
We thank the Fermilab staff and the technical staffs of the participating institutions for their vital contributions. This work was supported by the U.S. Department of Energy and National Science Foundation; the Italian Istituto Nazionale di Fisica Nucleare; the Ministry of Education, Culture, Sports, Science and Technology of Japan; the Natural Sciences and Engineering Research Council of Canada; the National Science Council of the Republic of China; the Swiss National Science Foundation; the A.P. Sloan Foundation; the Bundesministerium f\"ur Bildung und Forschung, Germany; the Korean World Class University Program, the National Research Foundation of Korea; the Science and Technology Facilities Council and the Royal Society, UK; the Institut National de Physique Nucleaire et Physique des Particules/CNRS; the Russian Foundation for Basic Research; the Ministerio de Ciencia e Innovaci\'{o}n, and Programa Consolider-Ingenio 2010, Spain; the Slovak R\&D Agency; and the Academy of Finland. 

\end{acknowledgments}
 %

\begin{thebibliography}{37}
\bibitem[{C-c()}]{C-conjugate}                                                   
\bibinfo{title}{{\rm Throughout this paper, $\mathit{C}$-conjugate modes are implied  
 and branching fractions indicate $\mathit{CP}$-averages.}}
\bibitem[{CDF()}]{CDF-coordinates}
\emph{\bibinfo{title}{{\em CDF II uses a cylindrical coordinate system in which
 $\phi$ is the azimuthal angle, $r$ is the radius from the nominal beam line,
 and $z$ points in the proton beam direction, with the origin at the center of
 the detector. The transverse plane is the plane perpendicular to the $z$
 axis.}}}
\bibitem[{Iso()}]{Isolation}
\bibinfo{title}{{Isolation is defined as
 $I_{B}=p_T(B)/(p_T(B)+\sum_ip_{Ti})$, where $p_T(B)$ is the transverse
 momentum of the $B$ candidate, and the sum runs over all other tracks within
 a cone of radius 1, in $\eta$-$\phi$ space around the $B$ flight-direction.}}
\end{thebibliography}
 %



\end{document}